\documentclass[aps,superscriptaddress,nofootinbib]{revtex4}
\usepackage{hyperref}
\usepackage{epsfig,rotating}
\usepackage{amsmath,amssymb}
\usepackage{dsfont}

\numberwithin{equation}{section}

\newcommand{\tm}{\theta_m}

\newcommand{\wtg}{\widetilde{\gamma}}

\newcommand{\ook}{\overline{\omega}(k)}

\newcommand{\be}{\begin{equation}}
\newcommand{\ee}{\end{equation}}
\newcommand{\bea}{\begin{eqnarray}}
\newcommand{\eea}{\end{eqnarray}}

\newcommand{\wk}{\overline{\omega}(k)}
\begin{document}

\title{Production of a sterile species via active-sterile mixing:\\ an exactly solvable model.}
\author{D. Boyanovsky}
\email{boyan@pitt.edu} \affiliation{Department of Physics and
Astronomy, University of Pittsburgh, Pittsburgh, Pennsylvania 15260,
USA}

\date{\today}

\begin{abstract}
The production of a sterile species via active-sterile mixing in a
thermal medium is studied in an   exactly solvable model. The
\emph{exact} time evolution of the sterile distribution function is
determined by the   dispersion relations and damping rates
$\Gamma_{1,2}$ for the quasiparticle modes. These depend on
  $\wtg = \Gamma_{aa}/2\Delta E$, with
$\Gamma_{aa}$ the interaction rate of the active species in absence
of mixing and $\Delta E$ the oscillation frequency in the medium
without damping. $\wtg \ll1,\wtg \gg 1$ describe the weak and strong
damping limits respectively.   For $\wtg\ll1$, $\Gamma_1 =
\Gamma_{aa}\cos^2\tm ; \Gamma_{2}=\Gamma_{aa}\sin^2\tm$ where $\tm$
is the mixing angle in the medium and the sterile distribution
function \emph{does not} obey a simple rate equation. For $\wtg \gg
1$, $\Gamma_1= \Gamma_{aa}$ and $\Gamma_2 = \Gamma_{aa}
\sin^22\tm/4\wtg^2$, is the sterile production rate. In this regime
sterile production  is suppressed and the oscillation frequency
\emph{vanishes} at an MSW resonance, with a breakdown of
adiabaticity. These are consequences of quantum Zeno suppression.
For active neutrinos with standard model interactions the strong
damping limit is \emph{only} available near an MSW resonance
\emph{if} $\sin2\theta \ll \alpha_w$ with $\theta$ the vacuum mixing
angle.  The  full set of quantum kinetic equations for sterile
production for arbitrary $\wtg$ are obtained from the quantum master
equation.  Cosmological  resonant sterile neutrino production is
quantum Zeno suppressed relieving potential uncertainties associated
with the QCD phase transition.

\end{abstract}

\pacs{98.80.Cq;14.60.Pq;11.10.Wx}

\maketitle

\section{Introduction}\label{sec:intro}
Sterile neutrinos, namely weak interaction singlets, are compelling
candidates to explain a host of cosmological and astrophysical
phenomena. They could be a suitable warm  dark matter
component\cite{dodelson,asaka,shi,kev1,hansen,kev2,kev3,kusenko,kou,dolgovrev,pastor,hannestad,biermann,michaDM},
  may also be relevant in the latest stages of stellar
  collapse\cite{raffeltSN,fuller},
primordial nucleosynthesis\cite{fuller2,fuller3}, and provide a
potential explanation for the anomalous velocity distributions of
pulsars\cite{segre,fullkus,kuse2}. Although sterile neutrinos are
ubiquitous in extensions of the standard
model\cite{book1,book2,book3,raffelt}, the MiniBooNE
collaboration\cite{miniboone} has recently reported   results in
contradiction with   those from LSND\cite{lsnd1,lsnd2} that
suggested a   sterile neutrino with $\Delta m^2 \sim
1~\textrm{eV}^2$ scale. Although the MiniBooNE results  hint at an
excess of events below $475~\mathrm{MeV}$ the analysis distinctly
excludes two neutrino appearance-only from $\nu_\mu \rightarrow
\nu_e$ oscillations  with a mass scale $\Delta m^2\sim
1~\textrm{eV}^2$, perhaps ruling out a  \emph{light} sterile
neutrino. However, a recent analysis\cite{malto} suggests that while
  $(3+1)$ schemes are strongly disfavoured, $(3+2)$ neutrino schemes
  provide a good fit to both the LSND and MiniBooNE data, including the low energy
  events, because of the possibility of CP violation in these schemes, although
  significant tension remains. These issues notwithstanding the MiniBooNE result does not
  constrain a heavier variety of sterile neutrinos such as those that could be suitable warm dark matter
  candidates with   masses in
the $\mathrm{keV}$
range\cite{dodelson,asaka,shi,kev1,hansen,kev2,kev3,kou,pastor,hannestad}.
Their   radiative decay   would contribute to the X-ray
background\cite{hansen,Xray,kou,boyarsky,hansen2} from which
constraints on their masses and mixing angles   may be
extracted\cite{kou,boyarsky,hansen2,kou2}.  It has also been
suggested that precision laboratory experiments   may be sensitive
to $\sim \textrm{keV}$ neutrinos\cite{shapolast}. Being weak
interaction singlets, sterile neutrinos can only be produced via
their mixing with an active species, hence any assessment of the
possibility of sterile neutrinos as dark matter candidates or their
role in supernovae must begin with understanding their production
mechanism. To be a suitable  dark matter candidate, two important
constraints must be satisfied: the correct abundance and a velocity
dispersion that restricts the free streaming length   to be
consistent with the constraints from  structure formation. Both
ingredients depend directly on the distribution function of the
sterile neutrinos, which in turn depend   on the  dynamics of
production and evolution until freeze-out.

Pioneering work on the  non-equilibrium dynamics of neutrinos in a
medium    was cast in terms of kinetic equations for a flavor
``matrix of densities''\cite{dolgov} or in terms of $2\times 2$
Bloch-type equations for flavor quantum mechanical
states\cite{stodolsky,enquist}. A general field theoretical approach
to neutrino mixing and kinetics was presented in \cite{sigl,raffkin}
(see also \cite{raffelt}), however  sterile neutrino production in
the early Universe is mostly studied in terms of simple
phenomenological rate
equations\cite{dodelson,kev1,cline,kainu,foot,dibari}, and numerical
studies \cite{kev1,dibari} rely on an approximate
semi-phenomenological approach\cite{kainu,foot}. A field theoretical
study of the hadronic contribution to the sterile production rate
\emph{near an MSW resonance}   has been reported in
ref.\cite{shapo}.

Understanding the  dynamics of oscillations, decoherence and damping
is of \emph{fundamental} and phenomenological importance not only in
neutrino cosmology   but also in the dynamics of neutral meson
mixing and CP violation\cite{cp,cp2,beuthe} and axion-photon mixing
in the presence of a magnetic field\cite{raffelt}, a phenomenon
whose interest has been  rekindled by the recent results from the
PVLAS collaboration\cite{pvlas}  (see the discussion in
ref.\cite{pvlas2}). As argued in\cite{dolokun} the spinorial nature
of neutrinos is inessential to describe the dynamics of   mixing and
decoherence   in a medium.

Recently we reported on a  study\cite{hobos} of mixing and
decoherence   in a theory of mesons that provides an accurate
  description of similar phenomena for mixed neutrinos. This
effective theory incorporates interactions that model the medium
effects associated with charge and neutral currents for neutrinos
and yields a   picture of the   dynamics   which is remarkably
general. The fermion nature of the distributions and Pauli blocking
effects can be simply accounted for in the final result\cite{hobos}.
This study implemented quantum field theory methods   to obtain the
non-equilibrium effective action for the ``neutrino'' degrees of
freedom. More recently this approach was extended to study the
production of sterile neutrinos both from the effective action as
well as from the correct quantum kinetic equations obtained directly
from the quantum master equation\cite{hoboyste}. The results
obtained in ref.\cite{hoboyste} clarify a host of important aspects,
such as the approach to equilibrium and  a detailed analysis of
quantum Zeno suppression when the decoherence time scale is shorter
than the oscillation time scale, thereby confirming previous results
obtained for   neutrinos with standard model interactions in
refs.\cite{hozeno,hochar}. The study in refs.\cite{hobos,hoboyste}
relied on integrating out the bath degrees of freedom, assumed to
remain in equilibrium, up to second order in a perturbative
expansion akin to an expansion in $G_F$ in the standard model. This
perturbative treatment restricted the analysis to the weak damping
regime in which the decoherence time scale is larger than the
oscillation time scale. In refs.\cite{hoboyste,hozeno} it was
pointed out   that a strong damping regime featuring the opposite
relation between these time scales could emerge near an MSW
resonance for small vacuum mixing angle consistent with constraints
from the X-ray background\cite{kou,boyarsky,hansen2,kou2}.

{\bf Motivation and goals:} A sound assessment of  sterile neutrinos
as warm dark matter candidates requires a reliable description of
the kinetics of production and evolution towards freeze-out. Strong
departure from equilibrium in the distribution function at
freeze-out could lead to significant changes in the abundance or
skewed velocity distributions that could affect the free streaming
lengths and structure formation\cite{fullshi}. In this article we
complement and extend a previous study\cite{hoboyste} on the
non-equilibrium production of a sterile species via active-sterile
mixing. While the previous study\cite{hozeno,hobos,hoboyste} focused
on the weak damping limit consistently with a perturbative expansion
in standard model interactions, this article studies an
\emph{exactly solvable} model that allows to explore systematically
the \emph{strong damping case} and to draw general conclusions on
the     production dynamics of a sterile species.

The model  incorporates all the relevant ingredients: active-sterile
mixing via a mass matrix which is off-diagonal in the flavor basis,
and the coupling of the active species to a continuum of degrees of
freedom which are taken as a thermal bath in equilibrium and
includes an index of refraction contribution which modifies the
mixing angles and dispersion relations in   the same manner as for
neutrinos propagating in a medium.

{\bf Summary of results:} The exact solution of the  Heisenberg
equations of  motion allows a complete investigation of the
non-equilibrium dynamics of production of the sterile species in the
weak and strong damping regimes and to analyze in detail   quantum
Zeno suppression. We   obtain the quantum master equation and from
it the complete set of kinetic equations that describe the
production and evolution of the active and sterile distribution
functions and coherences and reproduce the exact results. Our main
results are :

\begin{itemize}
\item{The exact solution of the Heisenberg (-Langevin) equations of
motion for one active and one sterile species yield two different
modes of propagation in the medium corresponding to quasiparticles
whose dispersion relations and damping rates (widths) depend on the
dimensionless ratio $\wtg = \Gamma_{aa}/2\Delta E$ with
$\Gamma_{aa}$ the active species interaction rate in absence of
mixing, and $\Delta E$ the oscillation frequency  in absence of
damping but including the index of refraction in the medium. The
weak and strong damping cases correspond to  $\wtg \ll 1$ and $\wtg
\gg 1$ respectively. The exact    distribution functions for the
active and sterile species are obtained, their time evolution is
completely determined by the  widths of these quasiparticles and the
oscillation frequency including corrections from the index of
refraction \emph{and damping}. }

\item{ The results in the weak damping regime $\wtg \ll 1$ coincide with
those obtained previously in refs.\cite{hobos,hoboyste,hozeno}: the
dispersion relations are akin to those of neutrinos in a medium with
an index of refraction and the damping rates are
$\Gamma_1=\Gamma_{aa} \cos^2\tm~;~\Gamma_2=\Gamma_{aa} \sin^2\tm$
where $\tm$ is the mixing angle in the medium. The generalized
active-sterile transition probability obtained from expectation
values of Heisenberg operators in the full quantum density matrix is
$\frac{\sin^22\theta_m}{4
}\left[e^{-\Gamma_1t}+e^{-\Gamma_2t}-2e^{-\frac{1}{2}(\Gamma_1+\Gamma_2)t}
\cos\left[\Delta E t\right]\right]$. The production of the sterile
species \emph{cannot} be described by a simple rate equation, since
the distribution function depends on the  time scales
$1/\Gamma_1,1/\Gamma_2,1/\Delta E$.}

\item{In the strong damping regime $\wtg \gg 1$ the oscillation
frequency \emph{vanishes at an MSW resonance} signaling a breakdown
of adiabaticity, and the widths of the quasiparticles become
$\Gamma_1 \sim \Gamma_{aa}$, $\Gamma_2 \sim \Gamma_{aa} \sin^2
2\tm/4\wtg^2$. To leading order in $1/\wtg$, the time evolution of
the sterile distribution function simplifies into a rate equation,
with the production rate given by $\Gamma_2 \sim \sin^22\theta_m
(\Delta E)^2/\Gamma_{aa}$ (see eqn. (\ref{npg2})). The
active-sterile transition probability is strongly suppressed $\sim
1/\wtg^2$. The vanishing of the oscillation frequency,   the
suppression of the transition probability and the production of the
sterile species are all manifestations of the \emph{quantum Zeno
effect} emerging in the strong damping limit. }

\item{For active neutrinos with standard model interactions it is shown that the strong
damping limit is only available near an MSW resonance for small
vacuum mixing angle $\theta$ satisfying the condition $\sin2\theta
\lesssim \alpha_w$ where  $\alpha_w$. This condition is likely
satisfied by the constraints on the vacuum mixing angle from the
X-ray background\cite{hansen,Xray} and entails that sterile neutrino
production is \emph{strongly suppressed  by the quantum Zeno effect
near an MSW resonance.} This suppression may relieve uncertainties
from the QCD phase transition for keV sterile neutrinos. }

\item{The quantum master equation for the reduced density matrix is
obtained under standard approximations.  From it the generalized
transition probability and the complete set of kinetic equations are
obtained valid in all regimes of damping. These reproduce the
results obtained from the exact treatment. Under simple
approximations the full set of kinetic equations is presented in the
form of quantum kinetic equations for a ``polarization vector''. The
complete set of kinetic equations (\ref{dotn11}-\ref{dotn12}) along
with the relations (\ref{Naoftfin},\ref{Nsoftfin}) provide a
complete description of the non-equilibrium evolution of the active
and sterile distributions and coherences.}

\end{itemize}

\section{The Model} \label{sec:model}

The main ingredients in the dynamics of the production of a sterile
species via active sterile mixing are: i) a mass matrix off diagonal
in the flavor basis which mixes the sterile and active species, ii)
the coupling of the active species to a bath in equilibrium. In the
standard model the bath degrees of freedom are quarks, leptons or
hadrons, these   equilibrate via strong or electromagnetic
interactions, hence can be taken to be in thermal equilibrium.

We propose  a simple exactly solvable model that includes all these
ingredients, it is  a generalization of a model for quantum Brownian
motion\cite{feyver,leggett} which has long served as a paradigm for
the study of quantum dissipative systems in condensed
matter\cite{qds} and quantum optics\cite{qobooks}. It consists of a
set of coordinates $\vec{q}$ that describe the ``system'' coupled to
a continuum of harmonic oscillators $Q_p$ that describe a thermal
bath in equilibrium. This simple model is generalized so that the
coordinates $ {q}_{a,s}$ stand for the active and sterile
``neutrinos'', these are mixed by off diagonal elements in a
frequency matrix but only the ``active'' coordinate couples to the
bath degrees of freedom. The motivation for studying this model stem
from the realization that the spinorial degrees of freedom are not
relevant to describe the non-equilibrium dynamics\cite{dolokun}, a
statement confirmed by previous studies of mixing, oscillations and
decoherence in a theory mesons\cite{hobos,hoboyste} which yields a
remarkably robust picture of the dynamics of neutrinos.

The Lagrangian for this model is

\bea L =   \frac{1}{2}\Bigg[  {\vec{\dot{q}}}^{~T} \cdot
\vec{\dot{q}}
   - \vec{ q}^{~T}~ \Big( k^2 \mathbb{I}+\mathbb{M}^2 +\mathbb{V}\Big)~\vec{q} \Bigg]+
  \frac{1}{2}\sum_p \Bigg[
 \dot{Q}^2_p-W^2_p Q^2_p\Bigg]+ q_a \sum_p C_p Q_p   \label{model}\eea
 where the   flavor vector is given by \be \vec{q} = \Bigg( \begin{array}{c}
                       q_a  \\
                       q_s
                     \end{array}\Bigg) \ee and  $k$ is a momentum
                     label, which is assumed but not included as an argument of $q_{a,s}$ for compact notation,
                     $\mathbb{I}$ is the $2\times 2$ identity matrix and \be \mathbb{M}^2  = \left( \begin{array}{cc}
                                   M^2_{aa} & M^2_{as} \\
                                   M^2_{as} & M^2_{ss} \\
                                 \end{array} \right)~~;~~\mathbb{V}   = \left( \begin{array}{cc}
                                   V_{aa}(k) & 0 \\
                                   0 & 0 \\
                                 \end{array} \right)\,.
                                 \label{matrices} \ee
The off diagonal elements of the mass matrix $\mathbb{M}$ lead to
active-sterile mixing and the matrix $\mathbb{V}$ models  a
``momentum dependent matter potential'' for the active species.

A sum over $k$ makes explicit the field theoretical nature of the
model, however just as in the case of neutrinos, we are interested
on the dynamics of a given $k$ mode in interaction with the ``bath''
degrees of freedom.

The correspondence with neutrinos is manifest by assuming that the
matter potential is obtained from one-loop charged and neutral
current contributions of $\mathcal{O}(G_F)$ from a background of
leptons, quarks or hadrons (or neutrinos in equilibrium) and
features a CP-odd term proportional to the lepton and baryon
asymmetries and a CP-even term that only depends on energy and
temperature\cite{notzold,bell}. The linear coupling of the active
species to the bath degrees of freedom with $C_p \propto G_F$ models
the charged current interaction, for example the coupling between
the electron neutrino and protons, neutrons and electrons in a
medium, $G_F
\overline{\psi}_P(C_V-C_A\gamma^5)\gamma^\mu\psi_N\overline{\psi}_e\gamma_\mu(1-\gamma^5)\nu_e$
(see    a similar description in\cite{raffelt,raffkin}). The label
$p$ will be taken to describe a continuum when the density of states
is introduced below. Obviously the model (\ref{model}) affords an
\emph{exact} solution and  yields   a remarkably general description
of the dynamics. The main ingredient is the coupling of a degree of
freedom to a \emph{continuum} of bath or environmental degrees of
freedom. Such coupling to a continuum is also at the heart of
particle-antiparticle oscillations in neutral meson systems
($K^0-\overline{K}^0; B^0-\overline{B}^0$) as described in
refs.\cite{beuthe,cp2}.  Other versions of this model, without
mixing   have   been studied with focus on the
  dynamics of equilibration\cite{boyalamo,hoboydavey}.

For vanishing matter potential $\mathbb{V}$ the flavor $q_{a,s}$ and
the mass coordinates $q_{1,2}$  are related by an orthogonal
transformation

\be   \left(\begin{array}{c}
      q_a \\
      q_s\\
    \end{array}\right) =  U(\theta) ~\Bigg(\begin{array}{c}
                                               q_1 \\
                                               q_2\\
                                             \end{array}\Bigg)~~;~~U(\theta)
                                             = \Bigg( \begin{array}{cc}
                           \cos\theta & \sin\theta \\
                           -\sin\theta & \cos\theta \\
                         \end{array} \Bigg) \label{trafo} \ee where
the orthogonal matrix $U(\theta)$ diagonalizes the mass matrix
$\mathds{M}^2$, namely

\be U^{-1}(\theta)\,\mathbb{M}^2 \, U(\theta) = \Bigg(
\begin{array}{cc}
                                         M^2_1 &0 \\
                                         0 & M^2_2 \\
                                       \end{array} \Bigg)
                                       \label{diagM} \ee and
                                       $\theta$ is the ``vacuum''
mixing angle in absence of the ``matter potential'' $\mathbb{V}$.

In the flavor basis the mass matrix $\mathbb{M} $ can be written in
terms of the vacuum mixing angle $\theta$ and the eigenvalues of the
mass matrix as

\be \mathbb{M}^2  = \overline{M}^{\,2}\,\mathds{1}+\frac{\delta
M^2}{2} \left(\begin{array}{cc}
                                                                -\cos 2\theta & \sin2\theta \\
                                                                \sin 2\theta & \cos 2\theta \\
                                                              \end{array}
\right) \label{massmatx2}\ee where we introduced

\be \overline{M}^{\,2} =\frac{1}{2}(M^2_1+M^2_2)~~;~~ \delta M^2 =
M^2_2-M^2_1 \label{MbarDelM}\,. \ee    The frequencies of the flavor
modes are determined by the diagonal entries of the matrix
$\mathbb{M}^2$ in the flavor basis, introducing \be \wk =
\sqrt{k^2+\overline{M}^2}\,, \label{wk}\ee   these  are given by \be
\omega_a(k) = \wk\Bigg[1-\frac{\delta
M^2}{2\,\wk^2}\cos2\theta\Bigg]^{\frac{1}{2}}~~;~~\omega_s(k) =
\wk\Bigg[1+\frac{\delta
M^2}{2\,\wk^2}\cos2\theta\Bigg]^{\frac{1}{2}} \label{flavfreqs}\ee

Focusing on the relevant case of ultrarelativistic neutrinos, we
anticipate that the \emph{only} approximation to be invoked is the
 one in which $\ook$ is larger than any other
energy scale. It is convenient to introduce \be \mathbb{K} \equiv
k^2\mathbb{I}+\mathbb{M}+\mathbb{V}=
\Bigg(\ook^2+\frac{V_{aa}}{2}\Bigg)\,\mathbb{I}+ \frac{\delta
M^2}{2}\, \Bigg[
                       \begin{array}{cc}
                         -\Big(\cos2\theta - \frac{V_{aa}}{\delta M^2}\Big) & \sin2\theta \\
                         \sin 2\theta & \Big(\cos2\theta - \frac{V_{aa}}{\delta M^2}\Big) \\
                       \end{array}
                     \Bigg]\,. \label{matx2}\ee

The exact solution will be presented in the Heisenberg picture, in
which  the density matrix is time independent and determined by its
initial value, which is assumed to be uncorrelated and of the form
\be\hat{\rho}(0) = \hat{\rho}_q \otimes \hat{\rho}_Q
\,.\label{DMat}\ee The   bath is taken to be in thermal equilibrium
with density matrix   $\hat{\rho}_Q = \mathrm{Tr}e^{-H_Q/T}$ where
$H_Q$ is the Hamiltonian for the sum of \emph{free} harmonic
oscillators of frequencies $W_p$.

                     The Heisenberg equations of
                     motion for the coordinates $q_{a,s},Q_p$ are
                     the following
\bea && \ddot{q}_{\alpha}+\mathbb{K}_{\alpha \beta}\, q_\beta =
\eta_\alpha~~;~~\alpha,\beta = a,s \label{qeqn} \\&&
\ddot{Q}_p+W^2_pQ_p = q_a C_p\,, \label{Qeqn}\eea where we have
introduced the flavor vector \be \vec{\eta} = \sum_p C_p Q_p ~\left(
               \begin{array}{c}
                 1 \\
                 0 \\
               \end{array}
             \right) \label{veceta}\,.\ee The solution of eqn
             (\ref{Qeqn}) is \be Q_p(t) = Q^{(0)}_p(t)+
             \frac{C_p}{W_p}\int^t_0
             \sin\big[W_p(t-t')\big]q_a(t')dt'
             \label{solQ}\,,\ee where \be Q^{(0)}_p(t) =
             \frac{1}{\sqrt{2W_p}}\Big[A_p\,e^{-iW_p t}+A^{\dagger}_p\,e^{iW_p t}\Big]
             \label{solQ0}\,,\ee is a solution of the homogeneous
             equation and $A_p,A^{\dagger}_p$ are free field
             annihilation and creation operators with the usual
             canonical commutation relations. The distribution function for the bath degrees of freedom is \be
             \mathrm{Tr}\,\hat{\rho}_Q\,  A^\dagger_p A_p =
             \frac{1}{e^{\frac{W_p}{T}}-1}=n(W_p)\label{nofp}\ee

             Introducing the
             solution  (\ref{solQ}) into (\ref{qeqn})
we find the Heisenberg-Langevin equations\cite{qobooks} \be
\label{HL} \ddot{q}_{\alpha}(t)+\mathbb{K}_{\alpha \beta}\,
q_\beta(t) + \int^t_0 \Sigma_{\alpha\beta}(t-t') q_\beta(t') =
\xi_\alpha (t) \ee where the self energy is diagonal in the flavor
basis and given by \be \label{SE} \Sigma_{\alpha\beta}(t-t') = -
\sum_p \frac{C^2_p}{W_p} \sin[W_p(t-t')] \left(
                                              \begin{array}{cc}
                                                1 & 0 \\
                                               0 & 0 \\
                                              \end{array}
                                            \right) \,.\ee
The \emph{stochastic} quantum noise is \be \vec{\xi}(t) =  \sum_p
C_p Q^{(0)}_p(t)~ \left(
               \begin{array}{c}
                1 \\
                 0 \\
               \end{array}
             \right) \label{noise}\,,\ee and we note that \be \mathrm{Tr}\hat{\rho}\, \vec{\xi}(t)
             =0\,.
             \label{xiav}\ee    The self energy $\Sigma$ is written in dispersive form  by passing
             to a continuum description of the bath degrees of freedom,  writing \be - \sum_p \frac{C^2_p}{W_p} \sin[W_p(t-t')] =  {i}
\int^\infty_{-\infty}\frac{d\omega}{\pi}
\textrm{Im}\Sigma_{aa}(\omega) e^{i\omega(t-t')} \label{SEdis}\ee
where the density of states
  \be \textrm{Im}\Sigma_{aa}(\omega) = \sum_p \frac{\pi
C^2_p}{2W_p}
\left[\delta(\omega-W_p)-\delta(\omega+W_p)\right]\label{disper}\ee
has the properties \be \textrm{Im}\Sigma_{aa}(-\omega)= -
\textrm{Im}\Sigma_{aa}(\omega)~~;~~ \textrm{Im}\Sigma_{aa}(\omega)>0
~~\textrm{for}~~ \omega >0\,. \label{SEprop}\ee The   density of
states  $\mathrm{Im}\Sigma_{aa}$ contains all of the relevant
information of the bath.  The Heisenberg-Langevin equation
(\ref{HL})is solved by Laplace transform, introduce  \be
\widetilde{q}_\alpha (s) = \int^\infty_0 e^{-st}q_\alpha(t)dt~~;
~~\textrm{etc}\,, \label{LT}\ee in terms of which the equation of
motion (\ref{HL}) becomes an algebraic equation \be \Bigg[s^2
\delta_{\alpha\beta}+\mathbb{K}_{\alpha\beta}+\widetilde{\Sigma}_{\alpha\beta}(s)\Bigg]\widetilde{q}_\beta(s)
= \dot{q}_\alpha(0)+s q_{\alpha}(0)+\widetilde{\xi}_\alpha(s)\,,
\label{LTE}\ee  where  in the flavor basis \be\widetilde{\Sigma}(s)
= -\frac{1}{\pi} \int^\infty_{-\infty}
\frac{\textrm{Im}\Sigma_{aa}(\omega')}{\omega'+is}d\omega' ~\left(
                                                    \begin{array}{cc}
                                                      1 & 0 \\
                                                      0 & 0 \\
                                                    \end{array}
                                                  \right)\,.\label{tilS}\ee
In what follows   we   need the analytic continuation of the
self-energy to real frequencies $s \rightarrow i\omega + 0^+$ \be
\widetilde{\Sigma}_{aa}(s=i\omega+0^+)=
\textrm{Re}\Sigma_{aa}(\omega) + i \, \textrm{Im}\Sigma_{aa}(\omega)
\label{conti}\ee with the dispersive relation  \be
\textrm{Re}\Sigma_{aa}(\omega) = -\frac{1}{\pi}\mathcal{P}
\int^\infty_{-\infty}
\frac{\textrm{Im}\Sigma_{aa}(\omega')}{\omega'-\omega}d\omega'
\,,\label{sigRE}\ee and $\mathcal{P}$ stands for the principal part.

 The
solution of eqn. (\ref{HL}) in real time is given by \be q_\alpha(t)
= \dot{G}_{\alpha\beta}(t)q_\beta(0)+
{G}_{\alpha\beta}(t)\dot{q}_\beta(0)+ \int^t_0
G_{\alpha\beta}(t')\xi_\beta(t-t')dt' \label{solt}\ee with \be
{G}_{\alpha\beta}(t) = \int_C \frac{ds}{2\pi i}~
\widetilde{G}_{\alpha\beta}(s)~ e^{st} \,. \ee The Laplace transform
of the propagator is given by \be \widetilde{G}(s) = \Bigg[s^2
\mathbb{I}+\mathbb{K}
+\widetilde{\Sigma}(s)\Bigg]^{-1}\label{tildeG}\ee and $C$ is the
Bromwich contour that runs parallel to the imaginary axis and to the
right of all the singularities of $\widetilde{G}$ in the complex
s-plane. It follows from eqns. (\ref{tildeG}) and (\ref{LTE}) that
the propagator matrix $G_{\alpha \beta}(t)$ is a homogeneous
solution of the equation of motion (\ref{HL}) with initial
conditions \be G_{\alpha \beta}(0)=0~;~\dot{G}_{\alpha \beta}(0)=1
\,. \label{Gini}\ee

It is convenient to introduce the following combinations \bea
\widetilde{\Delta}(s) & = & \frac{1}{\delta M^2}
\Big[\widetilde{\Sigma}_{aa}(s)+V_{aa}\Big] \label{Delta} \\
\widetilde{\rho}(s) & = &
\Bigg[\Big(\cos2\theta-\widetilde{\Delta}(s)\Big)^2+\sin^2 2\theta
\Bigg]\label{widerho}\eea and the matrix \be \mathbb{A}(s) =
\frac{1}{\widetilde{\rho}(s)}~\Bigg[
                                \begin{array}{cc}
                                  \cos2\theta-\widetilde{\Delta}(s) & -\sin 2\theta \\
                                  -\sin 2\theta & -\cos2\theta+\widetilde{\Delta}(s) \\
                                \end{array}
                              \Bigg]\,,\label{mtxA}\ee in terms of which we find \be \widetilde{G}(s) = \frac{1}{2}\,
\frac{\mathbb{I}+\mathbb{A}(s)}{s^2+\overline{\omega}^2(k)+\frac{\delta
M^2}{2}\big(\widetilde{\Delta}(s)-\widetilde{\rho}(s)\big)}
~+~\frac{1}{2}\,
\frac{\mathbb{I}-\mathbb{A}(s)}{s^2+\overline{\omega}^2(k)+
\frac{\delta
M^2}{2}\big(\widetilde{\Delta}(s)+\widetilde{\rho}(s)\big)}\,.\label{Gofs}\ee
Each term in this expression features poles   in the complex s-plane
near $s \approx \pm \, i \, \overline{\omega}(k)$ which are found by
first performing the analytic continuation $s\rightarrow i\omega +
0^+$ upon which the denominators in $\widetilde{G}(s)$ become \be
s^2+\overline{\omega}^2(k)+  \frac{\delta
M^2}{2}\big(\widetilde{\Delta}(s)\mp\widetilde{\rho}(s)\big)
\rightarrow -\omega^2+\overline{\omega}^2(k)+\frac{1}{2}
\Bigg[\textrm{Re}\Sigma_{aa}(\omega) + i \,
\textrm{Im}\Sigma_{aa}(\omega)+V_{aa} \Bigg]\mp \frac{\delta
M^2}{2}\rho(\omega) \ee where the analytic continuations are given
by\bea \rho(\omega) & = & \Bigg[\Big(\cos 2\theta -
\Delta_R(\omega)-i\Delta_I(\omega)\Big)^2+(\sin2\theta)^2
\Bigg]^{\frac{1}{2}}\label{rhomega}\\ \Delta_R(\omega) & = &
  \frac{\Big[\textrm{Re}\Sigma_{aa}(\omega)+V_{aa}
\Big]}{{\delta M^2}}~~;~~\Delta_I(\omega)  =
\frac{\textrm{Im}\Sigma_{aa}(\omega)}{\delta M^2}\,. \label{DRI}
\eea The complex poles describe \emph{quasiparticles}, the real part
determines their dispersion relation and the imaginary part their
damping rate in the medium.  At this stage it is convenient to
introduce the following variables \be \overline{\Delta}_R \equiv
\Delta_R(\overline{\omega}(k)) =
  \frac{\Big[ V_{aa}+ \textrm{Re}\Sigma_{aa}(\overline{\omega}(k))
\Big]}{\delta M^2}\ee   \be \widetilde{\gamma} \equiv
\frac{\Delta_I(\overline{\omega}(k))}{\rho_0} =
\frac{\textrm{Im}\Sigma_{aa}(\overline{\omega}(k))}{\delta
M^2\,\rho_0}\label{gamatil}\ee and write \be
\rho(\overline{\omega}(k)) = \rho_0\,r \,e^{-i\alpha} \ee where \bea
\rho_0  & = &  \Bigg[\Big(\cos 2\theta - \overline{\Delta}_R
\Big)^2+\Big(\sin2\theta\Big)^2\Bigg]^{\frac{1}{2}}\label{rho0}
\\ r & = & \Bigg[\Big(1 -
\widetilde{\gamma}^2 \Big)^2+\Big(2
\widetilde{\gamma}\cos2\theta_m\Big)^2\Bigg]^{\frac{1}{4}}\,,
\label{r} \\  \alpha & = & \frac{1}{2}~ \textrm{arctg} \Bigg[
\frac{2 \widetilde{\gamma}\cos2\theta_m}{ 1 - \widetilde{\gamma}^2
}\Bigg] \label{alpha}\eea  and the branch is chosen such that $0\leq
\textrm{arctg}\big[\cdots\big]\leq \pi$. The mixing angle in the
medium, $\tm$,  is defined by the relations \be \cos 2\theta_m =
\frac{\cos 2\theta - \overline{\Delta}_R }{\rho_0} ~~;~~
\sin2\theta_m = \frac{\sin 2\theta}{\rho_0} \,, \label{thetam} \ee
an MSW resonance in the medium occurs
whenever\cite{book1,book2,book3} \be \cos 2\theta =
\overline{\Delta}_R \,. \label{MSW}\ee
The \emph{only} approximations to be used are the following \be
\frac{\delta M^2}{\overline{\omega}(k)}\ll
1~;~\frac{\textrm{Re}\Sigma_{aa}(\omega)}{\overline{\omega}(k)}\ll1~;~
\frac{\textrm{Im}\Sigma_{aa}(\omega)}{\overline{\omega}(k)}\ll1
\label{onlyappx}\ee these are all consistent with the
ultrarelativistic limit, small radiative corrections and the narrow
width limit, all
approximations used in the case of neutrinos. 
 Using these  approximations   we find the following complex poles:
\begin{itemize}

\item{ The first term in (\ref{Gofs}) features complex poles at   \be \omega  = \pm \, \Omega_1 + i\frac{\Gamma_1}{2}
\label{pole1} \ee with \bea && \Omega_1  = \overline{\omega}(k)+
\frac{1}{4\overline{\omega}(k)}\Bigg[
\textrm{Re}\Sigma_{aa}(\overline{\omega}(k))+V_{aa}-\delta
M^2~\rho_0 \,r  \cos\alpha  \Bigg] \label{ome1}\\&&  \Gamma_1  =
   ~\frac{\Gamma_{aa}}{2}\Big[1+ \frac{ r
\sin\alpha}{\widetilde{\gamma}}\Big] \label{gama1c} \eea }

\item{  The second term in (\ref{Gofs}) features complex poles at    \be \omega  = \pm \, \Omega_2 + i\frac{\Gamma_2}{2}
\label{pole2} \ee with \bea && \Omega_2  = \overline{\omega}(k)+
\frac{1}{4\overline{\omega}(k)} \Bigg[
 \textrm{Re}\Sigma_{aa}(\overline{\omega}(k))+V_{aa}+\delta
M^2~\rho_0 \,r  \cos\alpha\Bigg] \label{ome2}\\&&  \Gamma_2   =
 ~\frac{\Gamma_{aa}}{2}\Big[1- \frac{ r
\sin\alpha}{\widetilde{\gamma}}\Big] \label{gama2c} \eea }

\end{itemize} where \be \Gamma_{aa} = \frac{
\textrm{Im}\Sigma_{aa}(\overline{\omega}(k))}{ \overline{\omega}(k)}
\label{Gamaaa}\ee is the interaction rate for the active species
\emph{in absence of mixing} in the limit $\overline{\omega}(k) \gg
\delta M^2$, which is of relevance for ultrarelativistic or nearly
degenerate neutrinos. In what follows we suppress the argument
$\overline{\omega}(k)$ in the quantities $\Delta_{R,I}$, etc.,  to
simplify notation.

Near the complex poles the analytic continuation
$\widetilde{G}(s=i\omega+0^+)$ features a Breit-Wigner form, and the
inverse Laplace transform can be performed by approximating the
analytic continuation by the   Breit-Wigner Lorentzian. We find

\bea G(t) = && \frac{e^{i\Omega_1t} ~ e^{-\frac{\Gamma_1}{2}
t}}{2i\Omega_1}\frac{1}{2}\Big[\mathbb{I}+\mathbb{T}\Big]
-\frac{e^{-i\Omega_1t} ~e^{-\frac{\Gamma_1}{2}
t}}{2i\Omega_1}\frac{1}{2}\Big[\mathbb{I}+\mathbb{T}^*\Big]+
\nonumber \\ && \frac{e^{i\Omega_2t} ~e^{-\frac{\Gamma_2}{2}
t}}{2i\Omega_2}\frac{1}{2}\Big[\mathbb{I}-\mathbb{T}\Big]
-\frac{e^{-i\Omega_2t} ~e^{-\frac{\Gamma_2}{2}
t}}{2i\Omega_2}\frac{1}{2}\Big[\mathbb{I}-\mathbb{T}^*\Big]\label{Goft}\eea
where we have neglected wave function renormalization (residues at
the poles)  and introduced the complex matrix \be \mathbb{T} =
\frac{e^{i\alpha}}{{r}} \left[
                              \begin{array}{cc}
                                \cos2\theta_m -i\widetilde{\gamma}  & -\sin 2\theta_m \\
                                -\sin 2\theta_m  & -\cos2\theta_m +i\widetilde{\gamma} \\
                              \end{array}
                            \right]\label{Tmtx} \ee where all quantities are
                            evaluated at $\omega
                            =\overline{\omega}(k)$ and used the approximations (\ref{onlyappx}). Inserting the result (\ref{Goft}) into the solution
(\ref{solt}) we obtain the complete solution for the time evolution
of the Heisenberg operators. The Breit-Wigner approximation leading
to exponential damping in (\ref{Goft})    is a Markovian
approximation\cite{qobooks}. The full solution requires the initial
conditions on the Heisenberg operators $q(0),\dot{q}(0)$, it is
convenient to expand these in a basis of creation and annihilation
operators of flavor states, \be q_\beta(0) =
\frac{1}{\sqrt{2\omega_\beta}}\Big[a_\beta (0)+
a^\dagger_\beta(0)\Big]~~;~~\dot{q}_\beta(0) = -i
\frac{\omega_\beta}{ \sqrt{2\omega_\beta}}\Big[a_\beta(0) -
a^\dagger_\beta(0)\Big]~~;~~\beta= a,s \label{q0}\ee where
$\omega_{a,s}$ are the frequencies associated with flavor
eigenstates given by eqn. (\ref{flavfreqs}). Under the validity of
the approximations (\ref{onlyappx}), we can approximate \be
\omega_a\sim \omega_s \sim \Omega_1\sim \Omega_2 \sim
\overline{\omega}(k) \label{freqappx}\ee leading to a simplified
form   \bea q_\alpha(t) \approx &&
\frac{1}{\sqrt{2\overline{\omega}(k)}}
\Bigg\{e^{-i\Omega_1t}~e^{-\frac{\Gamma_1}{2}t}\,\frac{1}{2}\Big[\mathbb{I}+\mathbb{T}^*\Big]+
e^{-i\Omega_2t}~e^{-\frac{\Gamma_2}{2}t}\,\frac{1}{2}\Big[\mathbb{I}-\mathbb{T}^*\Big]\Bigg\}_{\alpha\beta}a_\beta(0)
+ h.c. +  \nonumber \\ && \int^t_0
G_{\alpha\beta}(t')\xi_\beta(t-t')dt'\,. \label{qfinoft}\eea Under
the same approximations, we find the Heisenberg annihilation
operators at an arbitrary time from \be a_\alpha(t) =
\sqrt{\frac{\omega_\alpha}{2}}
\left[q_\alpha(t)-\frac{p_\alpha(t)}{i\omega_\alpha}\right]~~;~~p_\alpha(t)=\dot{q}_\alpha(t)\,,
\label{aoft}\ee  these are given by \bea a_\alpha(t) \approx &&
\Bigg\{e^{-i\Omega_1t}~e^{-\frac{\Gamma_1}{2}t}~\frac{1}{2}\Big[\mathbb{I}+\mathbb{T}^*\Big]+
e^{-i\Omega_2t}~e^{-\frac{\Gamma_2}{2}t}~\frac{1}{2}\Big[\mathbb{I}-\mathbb{T}^*\Big]\Bigg\}_{\alpha\beta}a_\beta(0)+
\nonumber \\&&  \sqrt{\frac{\overline{\omega}(k)}{2}}  \int^t_0
\left[G(t')+ \frac{i
\dot{G}(t')}{\overline{\omega}(k)}\right]_{\alpha\beta}\xi_\beta(t-t')
\label{aHoft}\eea where we have used the initial condition
$G_{\alpha\beta}(0)=0$ (see eqn. (\ref{Gini}))  and
(\ref{onlyappx}). Under these same approximations we find \be G(t')+
\frac{i \dot{G}(t')}{\overline{\omega}(k)} \simeq
\frac{i}{\overline{\omega}(k)}\Bigg\{ e^{-i\Omega_1
t'}~e^{-\frac{\Gamma_1}{2}t'} \frac{1}{2}
\Big[\mathbb{I}+\mathbb{T}^*\Big]+ e^{-i\Omega_2
t'}~e^{-\frac{\Gamma_2}{2}t'} \frac{1}{2}
\Big[\mathbb{I}-\mathbb{T}^*\Big]\Bigg\} \label{GdotG} \ee

\subsection{Transition probability}
The result (\ref{aHoft}) allows us to obtain the \emph{generalized}
transition probability from expectation values of these operators in
the initial density matrix. Denoting $\langle a (t) \rangle =
\mathrm{Tr}\hat{\rho}\,a(t) $ and using the result (\ref{xiav}) we
find \be \langle a_\alpha(t) \rangle =
\Bigg\{e^{-i\Omega_1t}~e^{-\frac{\Gamma_1}{2}t}\,\frac{1}{2}\Big[\mathbb{I}+\mathbb{T}^*\Big]+
e^{-i\Omega_2t}~e^{-\frac{\Gamma_2}{2}t}\,\frac{1}{2}\Big[\mathbb{I}-\mathbb{T}^*\Big]\Bigg\}_{\alpha\beta}\langle
a_\beta(0) \rangle \,.\label{aav}\ee  Consider an initial density
matrix that yields an initial non-vanishing expectation value for
the annihilation operator of the \emph{active} component, but a
vanishing expectation value for the sterile one, namely \be \langle
a_a(0) \rangle \neq 0~;~\langle a_s(0) \rangle = 0 \label{ainitp}\ee
From the form of the matrix $\mathbb{T}$ given by eqn. (\ref{Tmtx})
we find the \emph{generalized} active-sterile transition probability
\be \mathcal{P}_{a\rightarrow s}(t) = \Bigg|\frac{\langle a_s(t)
\rangle}{\langle a_a(0) \rangle}\Bigg|^2 =
\frac{\sin^22\theta_m}{4\,r^2}\left[e^{-\Gamma_1t}+e^{-\Gamma_2t}-2e^{-\frac{1}{2}(\Gamma_1+\Gamma_2)t}
\cos\left[(\Omega_2-\Omega_1)t\right]\right]\label{Pas} \ee where
\be \Omega_2-\Omega_1 =  \frac{\delta M^2\,\rho_0\,r
}{2\overline{\omega}(k)}\,     \cos\alpha ~~;~~\Gamma_1+\Gamma_2 =
\frac{\textrm{Im}\Sigma_{aa}(\overline{\omega}(k))}{\overline{\omega}(k)}=
\Gamma_{aa}\,.\label{freqdif}\ee and $\Gamma_{1,2}$ are given by
eqns. (\ref{gama1c},\ref{gama2c}). The expression (\ref{Pas}) is
similar to the transition probability for particle-antiparticle
mixing of neutral mesons\cite{cp,cp2}.

 The
oscillatory term is a result of the coherent interference between
the quasiparticle states in the medium and its exponential
suppression in (\ref{Pas}) identifies the \emph{decoherence} time
scale $\tau_{dec} = 2/(\Gamma_1+\Gamma_2) = 2/\Gamma_{aa}$.

\subsection{Weak and strong damping: quantum Zeno suppression}

The above expressions for the propagation frequencies and damping
rates of the quasiparticle excitations in the medium lead to
  two different cases: \bea \big|  \widetilde{\gamma}\big|
  & \ll &  1 \Rightarrow \textbf{weak~damping} \label{weakdamp}\\
  \big|
 \widetilde{\gamma}\big|
  &  \gtrsim &  1 \Rightarrow \textbf{strong~damping} \label{strongdamp}
  \eea These conditions can be written in a more illuminating
  manner, from the definitions (\ref{gamatil}) and (\ref{Gamaaa}) it
  follows that
   \be \widetilde{\gamma} = \frac{\Gamma_{aa}}{2\Delta E}\label{ratio}
   \ee where \be \Delta E = \frac{\delta
   M^2 \,\rho_0}{2\overline{\omega}(k)}  \label{delE}\ee is the
  \emph{oscillation frequency in the medium in absence of damping}, namely $\Delta E$ is given by
  $|\Omega_2-\Omega_1|$ setting  $\Delta_I=0$, i.e, the
  difference in the propagation frequencies only arising from the
  index of refraction in the medium. 
  The dimensionless quantity $\widetilde{\gamma}$ is the ratio
  between the \emph{oscillation time scale} $1/\Delta E$ and the
  \emph{decoherence time scale} $2/\Gamma_{aa}$. When
  $\widetilde{\gamma}\gg 1$ the environment induced decoherence occurs on
  time scales \emph{much shorter} than the oscillation scale and
  active-sterile oscillations are strongly suppressed. In the
  opposite limit $\widetilde{\gamma}\ll 1$ there are many
  oscillations before the environment induces decoherence. 

  The strong damping condition
  (\ref{strongdamp}) is then recognized with the condition for
  quantum Zeno suppression by scattering in a
  medium\cite{stodolsky,kev1}. It corresponds to the limit in which
  the active mean free path is shorter than the oscillation length and decoherence by the medium
  suppresses active-sterile oscillations.

  \subsubsection{Weak damping case: $\Big|\widetilde{\gamma}\Big|\ll 1$}

For weak damping  it follows that \be r   \approx   1 ~~;~~
\sin\alpha \approx   \widetilde{\gamma}\cos 2\theta_m
\label{alphawd}   \ee  and   the widths $\Gamma_{1,2}$  given by
(\ref{gama1c},\ref{gama2c}) become \be \Gamma_1 =
\Gamma_{aa}\cos^2\theta_m ~~;~~ \Gamma_2 = \Gamma_{aa}\sin^2\theta_m
\,. \label{gamawd}\ee    For the oscillation frequency we obtain \be
\Omega_2-\Omega_1 = \Delta E = \frac{\delta M^2\,\rho_0}{2\ook}
\label{wdosc}\ee and \be \mathbb{T} \simeq \left(
                              \begin{array}{cc}
                                \cos2\theta_m   & -\sin 2\theta_m \\
                                -\sin 2\theta_m  & -\cos2\theta_m   \\
                              \end{array}
                            \right) = U^{-1}(\theta_m)\left(
                              \begin{array}{cc}
                                1  & 0 \\
                                0   & -1  \\
                              \end{array}
                            \right)  U(\theta_m)\label{Twd}\ee where
$U(\theta)$ is the unitary matrix given by eqn. (\ref{trafo}).
Introducing the Heisenberg annihilation and creation operators
\emph{in the medium} as \be \left(
                              \begin{array}{c}
                                a_1(t)  \\
                                a_2(t) \\
                              \end{array}
                            \right) = U^{-1}(\theta_m) \left(
                              \begin{array}{c}
                                a_a(t)  \\
                                a_s(t) \\
                              \end{array}
                       \right) \label{Hopmed} \ee and similarly with the creation operators,
                        the time evolution (\ref{aav}) in the weakly
                       damped case yields \be \left(
                              \begin{array}{c}
                                \langle a_1(t)\rangle  \\
                               \langle a_2(t) \rangle \\
                              \end{array}
                            \right) = \left(
                                        \begin{array}{cc}
                                          e^{-i\Omega_1t}~e^{-\frac{\Gamma_1}{2}t} & 0 \\
                                          0 & e^{-i\Omega_2t}~e^{-\frac{\Gamma_2}{2}t} \\
                                        \end{array}
                                      \right)
                             \left(
                              \begin{array}{c}
                                \langle a_a(0) \rangle \\
                               \langle a_s(0) \rangle \\
                              \end{array}
                       \right) \,.\label{Hopmedoft} \ee Therefore,
                       in the weak damping regime, the Heisenberg
                       operators $a^\dagger_{1,2}\,,\,a_{1,2}$ create
                       and annihilate the \emph{in-medium states}
                       that propagate with frequencies
                       $\Omega_{1,2}$ and their ensemble averages damp out
                       with the widths $\Gamma_{1,2}$.
The active-sterile transition probability in this limit, is obtained
from eqn. (\ref{Pas}), and is given by \be \mathcal{P}_{a\rightarrow
s}(t)  = \Bigg|\frac{\langle a_s \rangle (t)}{\langle a_a \rangle
(0)}\Bigg|^2 = \frac{\sin^22\theta_m}{4
}\left[e^{-\Gamma_1t}+e^{-\Gamma_2t}-2e^{-\frac{1}{2}(\Gamma_1+\Gamma_2)t}
\cos\left[\Delta E t\right]\right]\,.\label{Paswdl}\ee In the weakly
damped case the decoherence time scale $\tau_{dec}=2/\Gamma_{aa}$ is
much larger than the oscillation time scale $1/\Delta E$, hence many
oscillations take place before the interaction with the environment
leads to decoherence.

                       These results reproduce those of
references\cite{hobos,hozeno,hoboyste} and confirm their generality
and applicability to the case of neutrinos with standard model
interactions studied in ref.\cite{hozeno}.

\subsubsection{Strong damping case: $\Big|\widetilde{\gamma} \Big| \gg 1 $}

The case of (very) strong damping  yields the following
simplifications: \bea    r^2  & \sim &   \widetilde{\gamma}^2 -1 + 2
\cos^2 2\theta_m
 \label{rsd}\\
  {r}\,\sin\alpha   & \sim  &  \widetilde{\gamma}~
\Big[1-\frac{\sin^22\tm}{2\widetilde{\gamma}^2}  \Big]  \,,
\label{alfasd} \eea leading to the damping rates \bea && \Gamma_1
\simeq
 \Gamma_{aa} \Bigg[1-\frac{\sin^22\tm}{4\wtg^2} \Bigg]\approx
\Gamma_{aa} \label{Gamma1sd}\\&& \Gamma_2 \simeq
 \Gamma_{aa}\,\frac{\sin^22\tm}{4\wtg^2}\,.\label{Gamma2sd}\eea  This is a remarkable
 result,   the quasiparticle width $\Gamma_2$  becomes \emph{vanishingly small}
in the strong damping regime, with important consequences for
production of the sterile species as seen below. Furthermore, the
oscillation frequency is found to be \be \Omega_2-\Omega_1 =
\frac{\delta M^2 \rho_0}{2\,\ook}
 \cos2\theta_m  = \Delta E \cos 2\tm \,, \label{oscfreqsd}\ee this is another
remarkable result in the strong damping regime: the oscillation
frequency \emph{vanishes} at the MSW resonance. It follows from
eqns.  (\ref{alpha}) and (\ref{ome1},\ref{ome2}) that the vanishing
of the oscillation frequency at an MSW resonance is an \emph{exact
result} for any $\wtg^2
>1$. This result implies that there is a degeneracy right at the
resonance, and unlike the quantum mechanical case in which there is
no level crossing, in presence of strong environmental damping the
two propagating states in the medium become \emph{degenerate} at the
resonance leading to  a breakdown of adiabaticity. Furthermore, in
this
  regime the transition probability (\ref{Pas})
is strongly  suppressed by the  factor $1/
\widetilde{\gamma}^2\ll1$, it is given by  \be
\mathcal{P}_{a\rightarrow s}(t) = \Bigg|\frac{\langle a_s(t)
\rangle}{\langle a_a(0) \rangle}\Bigg|^2 =
\frac{\sin^22\theta_m}{4\, \widetilde{\gamma}^2}
 \,\left[e^{-\Gamma_1t}+e^{-\Gamma_2t}-2e^{-\frac{1}{2}(\Gamma_1+\Gamma_2)t}
\cos\left[(\Omega_2-\Omega_1)t\right]\right]\label{Passd} \ee     In
the strong damping limit $\Delta \Omega =|\Omega_2-\Omega_1| \leq
\Delta E $ hence it follows that $\tau_{dec} \ll 1/\Delta\Omega$ and
the interference term is strongly damped out before one oscillation
takes place. This is the quantum Zeno effect in which the rapid
scattering in the medium prevents the build up of
coherence\cite{stodolsky}.

The vanishing of the oscillation frequency, the suppression of the
transition probability and $\Gamma_2$  in the strong damping case
are all manifestations of the quantum Zeno effect. Of particular
importance is the vanishing of the oscillation frequency at the MSW
resonance because this entails a breakdown of adiabaticity.

\section{Production of the sterile species}

The number of sterile particles is given \be N_s(t) = \langle
a^\dagger_s(t) a_s(t) \rangle \label{Nofs}\ee where the Heisenberg
operators are given by eqns. (\ref{aHoft},\ref{GdotG}) and the
expectation value is in the density matrix (\ref{DMat}). Let us
consider the case in which the initial density matrix
$\widehat{\rho}_q$ is diagonal \emph{in the flavor basis} with
initial populations  \be N_a(0)= \langle a^\dagger_a(0)a_a(0)\rangle
~~;~~ N_s(0)=\langle a^\dagger_s(0)a_s(0)\rangle
\,.\label{ininums}\ee Using the results    (\ref{aHoft},\ref{GdotG})
and the stochastic noise given by eqn. (\ref{noise},\ref{solQ0})
with the averages (\ref{nofp},\ref{xiav} ) we find \be N_s(t) =
\mathcal{P}_{a\rightarrow s}(t) N_a(0) + \mathcal{P}_{s\rightarrow
s}(t) N_s(0)+ N^\xi_s(t) \label{Nst}\ee where
$\mathcal{P}_{a\rightarrow s}(t)$ is the active-sterile transition
probability given by eqn. (\ref{Pas}), and  \bea
\mathcal{P}_{s\rightarrow s}(t)  & = &
\Big|e^{-i\Omega_1t}\,e^{-\frac{\Gamma_1}{2}t} f_-
+e^{-i\Omega_2t}\,e^{-\frac{\Gamma_2}{2}t} f_+ \Big|^2 \label{Pss}\\
f_{\pm} & = & \frac{1}{2}\Big(1\pm \frac{e^{-i\alpha}}{
{r}}\big(\cos2\tm + i\wtg\big)\Big)\,. \label{fpm}\eea  The
contribution $N^\xi_s(t)$ is completely determined by the
correlation function of the noise in the initial density matrix, it
is given by \be N^\xi_s(t)= \frac{\sin^22\theta_m}{4\,r^2} \int
\frac{d\omega}{\pi}
\frac{\mathrm{Im}\Sigma_{aa}(\omega)}{2\,\ook}\,n(\omega)
\Big|F_1(\omega;t)-F_2(\omega;t)\Big|^2 \label{Nxi}\ee where
$n(\omega)=[e^{\omega/T}-1]^{-1}$ and \be F_i (\omega;t) =
\frac{e^{-i(\Omega_i-\omega)t}\,e^{-\frac{\Gamma_i}{2}t}-1}{\omega_i-\omega
- \frac{\Gamma_i}{2}} ~~;~~ i=1,2 \,.\label{Fi}\ee The frequency
integral is carried out by approximating the functions
$F_i(\omega;t)$ as Breit-Wigner Lorentzians   near their complex
poles, the result is found to be \bea N^\xi_s(t) = &&
\frac{\sin^2\theta_m \cos^2\theta_m}{r^2}\Bigg\{
\frac{\Gamma_{aa}}{\Gamma_1}\,n(\Omega_1) \big(1-e^{-\Gamma_1t}\big)
+ \frac{\Gamma_{aa}}{\Gamma_2}\,n(\Omega_2)
\big(1-e^{-\Gamma_2t}\big) \nonumber \\ && -
e^{-\frac{1}{2}(\Gamma_1+\Gamma_2)t}\,
\frac{\Gamma_{aa}\big[n(\Omega_1) +n(\Omega_2)\big]}{
 \Big(\frac{\Gamma_{aa}}{2}\Big)^2+\Big(\Omega_2-\Omega_1\Big)^2   }
 \Bigg[\frac{\Gamma_{aa}}{2}\Big(1-\cos(\Omega_2-\Omega_1)t
 \Big)+ (\Omega_2-\Omega_1)\sin(\Omega_2-\Omega_1)t
  \Big)  \Bigg]
\Bigg\} \label{Nxifin} \eea The set of equations
(\ref{Nst},\ref{Pas}, \ref{Pss}) and (\ref{Nxifin}) completely
determine the time evolution of the sterile distribution function
$N_s(t)$.

\subsection{Weak and strong damping limits}

\subsubsection{Weak damping: $\big|\widetilde{\gamma}\big| \ll 1$}

In the weak damping limit the results above yield \bea r & \sim & 1
~;~ \sin\alpha \sim \mathcal{O}(\widetilde{\gamma})~;~ \cos\alpha \sim 1 \nonumber \\
\Gamma_1 & \sim & \Gamma_{aa} \cos^2\tm
~;~ \Gamma_2 \sim \Gamma_{aa} \sin^2\tm \nonumber\\
\Omega_2-\Omega_1 & \sim & \frac{\delta M^2 \,\rho_0}{2\ook} =\Delta
E \label{wdquans}\eea which lead to the following expression for the
number density of the sterile species, valid for an initial density
matrix diagonal in the flavor basis and with $N_s(0)=0\,,N_a(0)\neq
0 $, \bea N_s(t) = &&  N_a(0)\,\frac{ \sin^22\theta_m}{4
}\left[e^{-\Gamma_1t}+e^{-\Gamma_2t}-2e^{-\frac{\Gamma_{aa}}{2} t}
\cos\left[\Delta E t\right]\right]  \nonumber \\ && + \sin^2\theta_m
\,n(\Omega_1) \big(1-e^{-\Gamma_1t}\big) +
\cos^2\theta_m\,n(\Omega_2) \big(1-e^{-\Gamma_2t}\big)
+\mathcal{O}(\wtg) \label{wdNs}\eea this result reproduces those in
ref. \cite{hoboyste} for $N_s(0)=0$. We note that the production of
the sterile species \emph{cannot} be described in terms of a simple
rate equation in the weak damping case because it depends on several
different time scales.

\subsubsection{Strong damping limit: $\big|\wtg\big| \gg 1$}

While the strong damping limit $\big|\wtg\big|  \gtrsim 1$ must be
studied numerically, progress can be made in the \emph{very} strong
damping regime $\big|\wtg\big|  \gg 1$. It will be seen below that
this regime is relevant for sterile neutrinos near an MSW resonance.
In this regime the above results yield \bea r^2 & \sim &
\widetilde{\gamma}^2   \nonumber \\ \Gamma_1 & \sim &
 \Gamma_{aa}\nonumber
\\  \Gamma_2  & \sim &
 \Gamma_{aa} ~\frac{\sin^22\tm}{4\wtg^2}\nonumber \\  \Omega_2-\Omega_1 & \sim &  \frac{\delta M^2
\rho_0}{2\,\ook}  \cos2\theta_m  = \Delta E \cos2\tm \,.
\label{sdquans}\eea The coefficients \bea \frac{1}{2\,r^2} \,
\frac{\Gamma^2_{aa}}{
 \Big(\frac{\Gamma_{aa}}{2}\Big)^2+\Big(\Omega_2-\Omega_1\Big)^2   }
  & = &   \frac{2 }{\wtg^2+\cos^22\tm} \sim \frac{2}{\wtg^2}\ll 1  \nonumber
  \\ \frac{1}{r^2}\,
  \frac{\Gamma_{aa}(\Omega_2-\Omega_1)}{\Big(\frac{\Gamma_{aa}}{2}\Big)^2+\Big(\Omega_2-\Omega_1\Big)^2} &
  = &     \frac{2 \cos2\tm }{\wtg(\wtg^2+\cos^22\tm)}   \sim   \frac{1}{  \wtg ^3} \ll 1\,  \label{sdcoeffs} \eea
  therefore the second line in the noise contribution (\ref{Nxifin})
  becomes subleading.    Furthermore the ratios \bea \frac{\Gamma_{aa}}{r^2 \,\Gamma_1}
   & \sim &  \frac{1}{\wtg^2} \ll 1\nonumber
\\\frac{\Gamma_{aa}}{r^2 \,\Gamma_2} & \sim & \frac{4}{\sin^22\tm} \eea   therefore, only   the
term with $\Gamma_{aa}/\Gamma_2$    survives in the first line in
(\ref{Nxifin}). Since the transition probability in the first term
in eqn. (\ref{Nst}) $\mathcal{P}_{a\rightarrow s} \propto 1/\wtg^2$
(see eqn. (\ref{Pas})) this term  is also strongly suppressed,
therefore in the strong damping limit \be N_s(t)\sim N^\xi_s(t) \sim
n(\Omega_2) \big(1-e^{-\Gamma_2t}\big) \,. \label{Nxisdlim}\ee Hence
in this   limit the sterile population obeys a simple \emph{rate}
equation \be \frac{dN_s(t)}{dt} = \Gamma_2
\big[n(\Omega_2)-N_s(t)\big]\,, \label{rateNs}\ee however the
\emph{sterile production rate} is \be \Gamma_2= \Gamma_{aa}
~\frac{\sin^22\tm}{4\wtg^2} \ll \Gamma_{aa}\, \label{Nsrate}\ee
becoming vanishingly small in the strong damping case. We conclude
that \emph{sterile species production is strongly suppressed in the
strong damping case as a consequence of the quantum Zeno effect} .
The non-perturbative nature of this result is manifest by writing
\be \Gamma_2 =
 {\sin^22\tm}  \frac{(\Delta E)^2}{\Gamma_{aa}}\,. \label{npg2}\ee

We note that with $\wtg = \Gamma_{aa}/2\Delta E$ (see eqn.
(\ref{ratio}))  this result coincides with the effective rate in the
quantum Zeno limit $2\Delta E/\Gamma_{aa} \ll 1$ obtained in
reference\cite{foot} and implemented in the numerical study in
refs.\cite{kev1,dibari}. However, we argue below that in the case of
sterile neutrinos, the strong damping limit is \emph{only available}
near an MSW resonance, and far away from this resonance the
non-equilibrium dynamics corresponds to weak damping and the time
evolution of $N_s(t)$ \emph{cannot} be described by a simple rate
equation.

\section{Quantum master and kinetic equations}
Although we have obtained the time evolution of the distribution
function from the exact solution of the Heisenberg-Langevin
equations (under the approximation (\ref{freqappx})), within the
cosmological setting it is more convenient to obtain a set of
quantum kinetic equations for the distribution functions. This is
achieved by obtaining first the quantum master equation for the time
evolution of the reduced density matrix. In the case of neutrinos,
the index of refraction term $V_{aa}$ is of first order in $G_F$
(Fermi's effective weak coupling) while the self-energy
$\Sigma=\Sigma_R+i\Sigma_I$ is of second order. Furthermore the
study in the previous sections clearly shows that the contribution
of the real part of the self-energy yields a second order
renormalization of the index of refraction which can be simply
absorbed into a redefinition of $V_{aa}$. The most important aspect
of the second order self-energy correction arise from its imaginary
part, which yields the damping rates of the collective quasiparticle
excitations. The production of the sterile species is associated
with this imaginary part, and not the real part of the self-energy,
which only renormalizes the index of refraction in the medium.
Therefore it is convenient to include the index of refraction in the
``non-interacting'' part of the Hamiltonian  by first diagonalizing
the Hamiltonian for the system's degrees of freedom $\vec{q}$
corresponding to the first term in the Lagrangian (\ref{model}).
This is achieved by introducing the \emph{mass eigenstates} in the
medium with the index of refraction as follows. The matrix
$\mathbb{K}$ in eqn. (\ref{matx2}) can be written as \be \mathbb{K}
= \Bigg(k^2+\overline{M}^{\,2}+\frac{V_{aa}}{2}\Bigg)\,\mathbb{I}+
\frac{\delta M^2\,\rho_0}{2}\, \Bigg[
                       \begin{array}{cc}
                         - \cos2\theta_m   & \sin2\theta_m \\
                         \sin 2\theta_m &  \cos2\theta_m   \\
                       \end{array}
                     \Bigg]\,,  \label{nuK}\ee where  the expressions for $\rho_0$ and the
                     mixing angle in the medium are the same as
(\ref{rho0}, \ref{thetam})) but neglecting the \emph{second} order
correction $\mathrm{Re}\Sigma_{aa}$ to the index of refraction. The
diagonalization of the Hamiltonian is achieved via the unitary
transformation (\ref{trafo}) but in terms of the mixing angle in the
medium $\tm$ that includes the correction from the index of
refraction, namely \be \left(\begin{array}{c}
      q_a \\
      q_s\\
    \end{array}\right) =  U(\theta_m) ~\Bigg(\begin{array}{c}
                                               q_1 \\
                                               q_2\\
                                             \end{array}\Bigg)~~;~~U(\theta)
                                             = \Bigg( \begin{array}{cc}
                           \cos\theta_m & \sin\theta_m\\
                           -\sin\theta_m & \cos\theta_m \\
                         \end{array} \Bigg)\,. \label{nutrafo} \ee
                         Again to avoid proliferation of indices we
 refer to the coordinates that diagonalize the Hamiltonian with the index of refraction with the labels
 $1,2$, which now \emph{should not} be identified with those labeling the
 complex poles in section (\ref{sec:model}).

 Expanding $q_{1,2}$ and their canonical momenta $p_{1,2}$ in terms of Heisenberg annihilation and
 creation operators \be q_i = \frac{1}{\sqrt{2\omega_i}}\Big[a_i +
 a^\dagger_i]~~;~~p_i = -i\frac{\omega_i}{\sqrt{2\omega_i}}\Big[a_i -
 a^\dagger_i] \label{qps}\ee where the frequencies in the medium are \bea \omega_1  & \sim  &
\ook +\frac{V_{aa}}{4\,\ook} -\frac{\delta M^2\,\rho_0}{4\,\ook}  \nonumber \\
  \omega _2  & \sim &
 \ook+\frac{V_{aa}}{4\,\ook} +\frac{\delta M^2\,\rho_0}{4\,\ook} \,. \label{nufreqs}\eea
Under the approximation (\ref{onlyappx}) the active and sterile
annihilation (and creation) operators $a_{a,s}$ are related to
$a_{1,2}$ as \be a_a = \cos\tm a_1 + \sin\tm a_2  ~~;~~ a_s    =
\cos\tm a_2 - \sin\tm a_1 \,.\label{as}\ee

  The total system-bath Hamiltonian
 becomes $H= H_0 + H_I $ where \bea H_0 & = &  \sum_{i=1,2} a^\dagger_i a_i \, \omega_i+\sum_p \frac{1}{2}\Big[P^2_p+W^2_p Q^2_p
 \Big]  \label{H0}\\ H_I & = &
 (q_1\cos\tm+q_2\sin\tm)\sum_p C_pQ_p \,.\label{Hint} \eea The
 density matrix in the interaction picture of $H_0$ is    \be \widehat{\rho}_{i}(t) = e^{i{H}_0 t}e^{-i{H}
 t}\,
\widehat{\rho}(0)\,e^{ i{H}  t}e^{-i{H}_0 t}\label{rhoip}\ee where
$\widehat{\rho}(0)$ is given by eqn. (\ref{DMat}). The equation of
motion of the density matrix in the interaction picture is  \be
\frac{d\widehat{\rho}_{i}(t)}{dt} =
-i\left[H_{I}(t),\widehat{\rho}_{i}(t)\right] \label{eqrhoip}\ee
with $H_I(t) =e^{i {H}_0 t} H_I e^{-i {H}_0 t}$ is the interaction
Hamiltonian in the interaction picture of $ {H}_0$. Iteration of
this equation up to second order in the interaction
yields\cite{qobooks} \be \frac{d\widehat{\rho}_{i}(t)}{dt} =
-i\left[H_{I}(t),\widehat{\rho}_{i}(0)\right]- \int^t_0 dt'\left[
 H_I(t),\left[H_I(t'),\widehat{\rho}_{i}(t')\right]\right]+\cdots
\label{2ndord}\ee

The \emph{reduced} density matrix for the system's variables $q$ is
obtained from the total density matrix by tracing over the bath
degrees of freedom $Q_p$,  which are assumed to remain in
equilibrium\cite{qobooks}. The following standard approximations are
invoked\cite{qobooks}:
   \textbf{a): factorization:}  the total density matrix is assumed
to factorize \be\widehat{\rho}_i(t)=
\rho_{q,i}(t)\otimes\rho_Q(0)\label{fact}\ee where it is assumed
that the bath remains in equilibrium.    \textbf{b): Markovian
approximation:} the memory of the evolution is neglected and in the
double commutator in (\ref{2ndord}) $\widehat{\rho}_i(t')$ is
replaced by $\widehat{\rho}_i(t)$ and taken out of the
integral\cite{qobooks}.
    Taking the trace over the bath degrees of freedom
yields the quantum master equation for the reduced density matrix,
\be \frac{d {\rho}_{R}(t)}{dt} =  - \int^t_0
dt'{\mathrm{Tr}}\rho_Q\big\{\left[
 H_I(t),\left[H_I(t'),\widehat{\rho}_{i}(t)\right]\right]\big\}+\cdots
\label{2ndordred}\ee where the first term has vanished because $Tr_Q
\rho_Q(0) Q^{(0)}_p(t) =0$ since $Q^{(0)}_p(t)$ is a free harmonic
oscillator in the interaction picture of $H_0$ (see eqn.
(\ref{solQ0})).  The trace over $Q$ in the double commutator
requires the following ingredients \bea \sum_{p,p'}\frac{ C_p
C_{p'}}{\sqrt{4W_p W_{p'}}}\, \mathrm{Tr}\rho_Q(0)
Q^{(0)}_p(t)Q^{(0)}_{p'}(t') & = & \sum_p
\frac{C^2_p}{2W_p}\Big[(1+n(W_p))\,e^{-iW_p(t-t')}+n(W_p)\,e^{iW_p(t-t')}\Big]
\nonumber \\ & = & \int \frac{d\omega}{\pi}
\mathrm{Im}\Sigma_{aa}(\omega)(1+n(\omega))\,e^{-i\omega(t-t')}
\label{gplus}\eea \bea \sum_{p,p'}\frac{ C_p C_{p'}}{\sqrt{4W_p
W_{p'}}}\, \mathrm{Tr}\rho_Q(0) Q^{(0)}_{p'}(t')Q^{(0)}_p(t) & = &
\sum_p
\frac{C^2_p}{2W_p}\Big[(1+n(W_p))\,e^{-iW_p(t'-t)}+n(W_p)\,e^{iW_p(t'-t)}\Big]\nonumber\\
 & = & \int \frac{d\omega}{\pi} \mathrm{Im}\Sigma_{aa}(\omega)\, n(\omega)
\,e^{-i\omega(t-t')} \label{gmin} \eea where the interaction picture
operators $Q^{(0)}(t)$ are given by eqn. (\ref{solQ0}) and we have
used eqns. (\ref{disper},\ref{SEprop}).

Several standard approximations are invoked: terms that feature
rapidly varying phases of the form $a^\dagger_i a^\dagger_j \,e^{i
(\omega_i+\omega_j)t}$ and $a_i a_j e^{ -i(\omega_i+\omega_j)t}$ are
averaged out in time leading to their cancellation, in the quantum
optics literature this is known as the ``rotating wave
approximation''\cite{qobooks}, similar terms are discarded in the
kinetic approach in ref.\cite{raffkin,raffelt}. The time integrals
are evaluated in the  Weisskopf-Wigner
approximation\cite{qobooks,hoboyste}. Finally we also invoke the
ultrarelativistic approximation $\omega_1 \sim \omega_2 \sim \ook$.
Neglecting the second order energy shift  (see eqn. (\ref{sigRE})),
the final result for the quantum master equation is given by \be
\frac{d\rho_{R}(t)}{dt} = -\frac{\Gamma_{aa}}{2} \Bigg\{\cos^2\tm
\mathcal{L}_{11}[\rho_R]+ \sin^2\tm \mathcal{L}_{22}[\rho_R]+
\frac{1}{2}\sin2\tm
\Big(\mathcal{L}_{12}[\rho_R]+\mathcal{L}_{21}[\rho_R] \Big)
\Bigg\}\label{qme}\ee where $\mathcal{L}_{ij}[\rho_R]$ are the
Lindblad operators\cite{qobooks} \bea \mathcal{L}_{ij}[\rho_R] = &&
\big(1+n(\omega_i)\big)\Big[\rho_R a^\dagger_i a_j+ a^\dagger_j a_i
\rho_R -  a_i \rho_R a^\dagger_j- a_j \rho_R a^\dagger_i \Big]
\nonumber \\ && + n(\omega_i) \Big[\rho_R a_i a^\dagger_j+ a _j
a^\dagger_i \rho_R - a^\dagger_i \rho_R a_j - a^\dagger_j \rho_R a_i
\Big] \label{Lij}\eea In these expressions, the annihilation and
creation operators carry the time dependence in the interaction
picture, namely \be a^\dagger_i(t)= a^\dagger_i(0)\,e^{i\omega_i
t}~~;~~a_i(t)= a_i(0)\,e^{-i\omega_i t}\,. \label{ipas}\ee   The
trace of the reduced density matrix is automatically conserved in
time as a consequence of unitary time evolution of the full density
matrix. Denoting the expectation value of any interaction picture
operator $A(t)$ in the reduced density matrix by \be \langle A
\rangle (t) = \mathrm{Tr}\rho_R(t) A(t) \,,\label{aveR}\ee  we
obtain the following equations for the expectation values of the
annihilation operators \be  \frac{d}{dt} \left(
                                                        \begin{array}{c}
                                                          \langle a_1 \rangle(t) \\
                                                          \langle a_2 \rangle(t) \\
                                                        \end{array}
                                                      \right)=
\left(
  \begin{array}{cc}
     -i\omega_1 -
\frac{\Gamma_{aa}}{2}\cos^2\tm  & -\frac{\Gamma_{aa}}{4}\sin2\tm  \\
   -\frac{\Gamma_{aa}}{4}\sin2\tm  &  -i\omega_2 - \frac{\Gamma_{aa}}{2}\sin^2\tm  \\
  \end{array}
\right) \left(
                                                        \begin{array}{c}
                                                          \langle a_1 \rangle(t) \\
                                                          \langle a_2 \rangle(t) \\
                                                        \end{array}
                                                      \right)
                                                      \label{amtx}\ee

The eigenvalues of the matrix in eqn. (\ref{amtx}) are found to be $
-i\widetilde{\Omega}_{1,2}-\Gamma_{1,2}/2$ where
$\widetilde{\Omega}_{1,2}$ are obtained  from eqns.
(\ref{ome1},\ref{ome2})  by  setting the second order contribution
to the energy shift $\mathrm{Re}\Sigma_{aa}=0$, and $\Gamma_{1,2}$
are \emph{precisely} given by eqns.(\ref{gama1c},\ref{gama2c}) but
again setting $\mathrm{Re}\Sigma_{aa}=0$ in  $\rho_0$, which of
course is a consequence of having neglected the second order energy
shifts (real part of the self energy) in the quantum master
equation. It is a straightforward exercise to obtain the (complex)
eigenvectors of the matrix (\ref{amtx}) and to write   $\langle
a_{a,s} \rangle$ in terms of these through the relation (\ref{as}).
Fixing the initial values of the corresponding eigenvectors to yield
the initial values $\langle a_a \rangle (0) \neq 0; \langle a_s
\rangle (0)  =0$ we find \be \mathcal{P}_{a\rightarrow s}(t)  =
\Bigg|\frac{\langle a_s \rangle (t)}{\langle a_a \rangle
(0)}\Bigg|^2 =
\frac{\sin^22\theta_m}{4\,r^2}\left[e^{-\Gamma_1t}+e^{-\Gamma_2t}-2e^{-\frac{1}{2}(\Gamma_1+\Gamma_2)t}
\cos\left[(\widetilde{\Omega}_2-\widetilde{\Omega}_1)t\right]\right]\label{Pasqme}\ee
which is the same as the transition probability (\ref{Pas}) but
neglecting the second order correction from
$\mathrm{Re}\Sigma_{aa}$. These results clearly show that the
quantum master equation (\ref{qme}) correctly describes the
non-equilibrium dynamics \emph{including the strong damping regime},
  the only difference with the exact result being   that the
second order energy shift $\mathrm{Re}\Sigma_{aa}$ is neglected .
The quantum master equation (\ref{qme}) is exactly the same as the
one obtained in ref.\cite{hoboyste}.

We now introduce the distribution functions \be n_{ij} =
\mathrm{Tr}\rho_R(t) a^\dagger_i(t) a_j(t)\,, \label{nij}\ee the
diagonal components describe the population of the \emph{in medium}
states, and the off-diagonal components the
coherences\cite{qobooks}. Accounting for the free field time
dependence of the operators $a^\dagger,a$ in the interaction
picture, we find the following kinetic equations for the
distribution functions \bea \dot{n}_{11} & = &  -\Gamma_{aa} \Big\{
\cos^2\tm \big(n_{11}-n(\omega_1)\big) + \frac{\sin2\tm}{4}
\big(n_{12}+n^*_{12}\big) \Big\} \label{dotn11}
\\\dot{n}_{22}  & = & -\Gamma_{aa} \Big\{ \sin^2\tm
\big(n_{22}-n(\omega_2)\big) + \frac{\sin2\tm}{4}
\big(n_{12}+n^*_{12}\big) \Big\} \label{dotn22}\\\dot{n}_{12} & = &
-i\Big(\omega_2-\omega_1\Big)n_{12}-\frac{\Gamma_{aa}}{2}\Bigg[n_{12}+\frac{\sin2\tm}{2}\big(n_{11}+n_{22}
-n(\omega_1)-n(\omega_2)\big)\Bigg]\label{dotn12}\eea where
$n(\omega_i)$ are the equilibrium distribution functions. In terms
of the $n_{ij}(t)$ we obtain the time evolution of the active and
sterile distribution functions via the relation (\ref{as}), namely
\bea N_a(t) &  = &  \cos^2\tm n_{11}(t) + \sin^2\tm n_{22}(t) +
\frac{1}{2} \sin2\tm \big(n_{12}(t)+n^*_{12}(t)\big)
\label{Naoftfin}\\N_s(t) &  = & \sin^2\tm n_{11}(t) + \cos^2\tm
n_{22}(t) - \frac{1}{2} \sin2\tm \big(n_{12}(t)+n^*_{12}(t)\big)\,.
\label{Nsoftfin} \eea The weak damping limit can be studied in a
perturbative expansion in $\wtg \ll 1$ by considering the terms
$n_{12},n^*_{12}$ in equations (\ref{dotn11},\ref{dotn22}) and the
terms $n_{ii}-n(\omega_i);i=1,2$ in equation (\ref{dotn12}) as
perturbations. This study was carried out in ref.\cite{hoboyste} and
reproduces the result eqn. (\ref{wdNs}) for the sterile population.
Therefore the set of quantum kinetic equations
(\ref{dotn11}-\ref{dotn12}) reproduce the exact results
  both in the weak and strong damping cases.

We can now establish a correspondence with the quantum kinetic
equation often quoted in the
literature\cite{mckellar,raffelt,foot,dibari,wong} by introducing
the following ``polarization vector''\cite{bohot} \bea P_0(t) & = &
\langle a^\dagger_a a_a + a^\dagger_s a_s  \rangle(t) = N_a( t)+N_s(
t)\label{P0}\\P_x( t) & = & \langle a^\dagger_a a_s + a^\dagger_s
a_a  \rangle (t) \label{Px}\\P_y( t) & = & -i\langle a^\dagger_a a_s
-a^\dagger_s a_a \rangle (t) \label{Py}\\P_z( t) & = & \langle
a^\dagger_a a_a - a^\dagger_s a_s  \rangle (t) = N_a( t)-N_s( t)
\label{Pz}\eea where the creation and annihilation operators for the
active and sterile fields are related to those that create and
annihilate the propagating modes in the medium $1,2$ by eqn.
(\ref{as}), and the angular brackets denote expectation values in
the reduced density matrix $\rho_{R}$ which obeys the quantum master
equation (\ref{qme}). In terms of the population and coherences
$n_{ij}$ the elements of the polarization vector are given by

\bea P_0 & = & n_{11}+n_{22} \label{P01} \\
P_x & = & -\sin2\tm \Big(n_{11}-n_{22}\Big)+\cos2\tm \big(n_{12}+n^*_{12} \big)\label{Px1}\\
P_y & = & - i\big(n_{12}- n^*_{12} \big) \label{Py1}\\P_z & = &
\cos2\tm \Big(n_{11}-n_{22}\Big)+ \sin2\tm \big(n_{12}+n^*_{12}
\big)\,. \label{Pz1}\eea Using the quantum kinetic equations
(\ref{dotn11}-\ref{dotn12}) we find \be \frac{dP_0}{dt} =
-\frac{\Gamma_{aa}}{2}P_z -
\frac{\Gamma_{aa}}{2}\Bigg[\Big(n_{11}-n(\omega_1)\Big)+
\Big(n_{22}-n(\omega_2)\Big) \Bigg]+\frac{\Gamma_{aa}}{2}\cos2\tm
\Big(n(\omega_1)-n(\omega_2)\Big)\label{dotP0}\ee \be
\frac{dP_x}{dt} = -i(\omega_2-\omega_1) \cos2\tm
\big(n_{12}-n^*_{12}) \big) - \frac{\Gamma_{aa}}{2}P_x -
\frac{\Gamma_{aa}}{2}\sin2\tm \Big(n(\omega_1)-n(\omega_2)\Big)
\label{dotPx}\ee \be \frac{dP_y}{dt} =
-(\omega_2-\omega_1)\big(n_{12}+n^*_{12}\big)
-\frac{\Gamma_{aa}}{2}P_y \label{dotPy}\ee \be \frac{dP_z}{dt} =
-i(\omega_2-\omega_1) \sin2\tm \big(n_{12}-n^*_{12})
-\frac{\Gamma_{aa}}{2}P_z -
\frac{\Gamma_{aa}}{2}\Bigg[\Big(n_{11}-n(\omega_1)\Big)+
\Big(n_{22}-n(\omega_2)\Big) \Bigg] \label{dotPz}\ee

Under the approximation $\omega_1\sim \omega_2 \sim \ook$ we can
take \be \Big(n(\omega_1)-n(\omega_2)\Big) \sim 0 \,,
\label{eqmass}\ee and neglect the last terms in eqns.
(\ref{dotP0},\ref{dotPx}). Introducing the vector $\vec{V}$ with
components \be \vec{V} = (\omega_2-\omega_1)~\Big(\sin 2\tm, 0 ,
-\cos2\tm\Big) \label{VecV}\ee   we find the following equations of
motion for the polarization vector \be \frac{d\vec{P}}{dt} =
\vec{V}\times\vec{P} - \frac{\Gamma_{aa}}{2}\Big(P_x \hat{x}+P_y
\hat{y}\Big)+ \frac{dP_0}{dt}\hat{z}\,. \label{QKEpol}\ee This
equation is exactly of the form \be \frac{d\vec{P}}{dt} =
\vec{V}\times\vec{P} - D \vec{P}_T + \frac{dP_0}{dt}\hat{z}
\label{QKEpol2}\ee often used in the
literature\cite{stodolsky,mckellar,wong,foot,dibari}, where \be  D =
\frac{\Gamma_{aa}}{2} ~~;~~ \vec{P}_T   =   \Big(P_x \hat{x}+P_y
\hat{y}\Big) \,.\label{DandPT}\ee

Therefore the quantum kinetic equation for the polarization vector
(\ref{QKEpol}) is \emph{equivalent} to the full set of quantum
kinetic equations (\ref{dotn11}-\ref{dotn12}).

However it must be highlighted that the set of equations
(\ref{QKEpol},\ref{QKEpol2}) is \emph{not closed} because it must
input the time evolution of $P_0$ which is obtained from the full
set of kinetic equations (\ref{dotn11}-\ref{dotn12}).

Often the last term in (\ref{QKEpol2}) ($\dot{P}_0$) is omitted,
however, such omission is   not warranted, since it follows from the
definition of $P_0$, eqn. (\ref{P01}) and eqns
(\ref{Naoftfin},\ref{Nsoftfin}), that \be P_0 = N_a(t)+N_s(t)\,, \ee
therefore $\dot{P}_0$    vanishes \emph{only} when both the active
and the sterile species have reached equilibrium. Thus we advocate
that the set of kinetic equations (\ref{dotn11}-\ref{dotn12})
combined with the relations (\ref{Naoftfin},\ref{Nsoftfin}) provide
a complete description of active and sterile production.

\section{Consequences for cosmological production of sterile neutrinos.}

The results obtained above can be straightforwardly adapted to the
case of neutrinos by replacing the equilibrium distributions
$n(\Omega_{1,2})$ by the Fermi-Dirac distributions in the
ultrarelativistic limit and the matter potential from forward
scattering in the medium.

While in general $\wtg$, $\Gamma_{1,2}$ and $\Omega_{1,2}$ depend on
the details of the interactions, masses and vacuum mixing angles, an
assessment of the consequences of the results obtained above on
cosmological sterile neutrino production can be obtained for an
active neutrino with standard model interactions. In this case the
matter potential for temperatures features a CP-odd contribution
proportional to the lepton and baryon asymmetries, and a CP-even
contribution that depends solely on momentum and temperature. In the
ultrarelativistic limit with $\ook\sim k$ the matter potential for
neutrinos  is given by\cite{notzold,bell,boyhohec}, \be V_{aa} =
\frac{4\sqrt{2}\xi(3)}{\pi^2} G_F k T^3    \left[L-A
\frac{Tk}{M^2_W} \right]   \label{matpotsm} \ee  where $L$ is
proportional to the lepton and baryon asymmetries and $A \sim
10$\cite{notzold,bell}, for antineutrinos $L\rightarrow -L$. The
active neutrino interaction rate (neglecting contributions from the
lepton and baryon asymmetries) is given by
\cite{notzold,bell,raffelt,cline,kainu} \be \Gamma_{aa} \sim G^2_F
T^4 k  \,.\label{smgama} \ee For $keV$ sterile neutrinos an MSW
resonance is available only for
 $L \gg Tk/M^2_W$ when the first term in the bracket in
 (\ref{matpotsm}) dominates\cite{bell,dibari,kev1,kev2}, while no
 resonance is available when the second term dominates.  We will
 analyze separately the two different cases \bea  L & \ll &
 \frac{T^2}{M^2_W}   \label{smallL}\\  L & \gg & \frac{T^2}{M^2_W}    \label{largeL}\eea  where we have taken
 $k \sim T$. In the first case no MSW resonance is possible for
 $keV$ sterile neutrinos, whereas such resonance is possible in the
 second case\cite{bell,dibari,kev1,kev2}.

\begin{itemize}

\item{{\bf High temperature limit:}} At high temperature above the MSW resonance for $V_{aa}\gg \delta
M^2 $ and neglecting the second order correction to the matter
potential ($\textrm{Re}\Sigma$), \be \rho_0 \sim
\frac{V_{aa}}{\delta M^2}\,.  \ee For $L \ll T^2/M^2_W$  \be
\frac{\delta M^2 \rho_0}{\ook} \sim \frac{G_F \,T^5}{M^2_W} \ee and
the ratio \be \wtg=  \Bigg|\frac{\Gamma_{aa}}{\frac{\delta
M^2}{\ook} \rho_0}\Bigg| \sim G_F M^2_W \sim \alpha_w \ll 1 \ee
where $\alpha_w$ is the standard model ``fine structure constant''.
For $L\gg T^2/M^2_W$  a similar analysis yields \be \wtg \sim G_F
M^2_W \left(\frac{T^2}{LM^2_W}\right) \sim \alpha_w
\left(\frac{T^2}{LM^2_W}\right) \ll1 \,. \ee

\item{\bf Low temperature limit:} In the   low temperature regime for $V_{aa}\ll \delta
M^2$,   $\rho_0 \sim 1$ and $\wtg$ becomes \be
\Big|\frac{\textrm{Im}\Sigma_{aa}}{\delta M^2 \rho_0} \Big| \sim
\Big|\frac{\textrm{Im}\Sigma_{aa}}{\delta M^2 } \Big| \ee however in
perturbation theory $V_{aa}\gg \textrm{Im}\Sigma $ since $V_{aa}$ is
of $\mathcal{O}(G_F)$ and $\textrm{Im}\Sigma_{aa} \sim
\mathcal{O}(G^2_F)$. Therefore since in this regime \be \delta M^2
\gg V_{aa}\gg \textrm{Im}\Sigma_{aa} \Rightarrow
\Big|\frac{\textrm{Im}\Sigma_{aa}}{\delta M^2 } \Big| \ll 1 \ee The
conclusion of this analysis is that \emph{far away from an MSW
resonance, either in the high or low temperature limit damping is
weak}, namely at high or low temperature away from the MSW resonance
\be \wtg=\frac{\Gamma_{aa}}{2\Delta E}\ll 1\,. \label{weak}\ee
Therefore the strong damping condition may \emph{only} be fulfilled
near an MSW resonance $ \theta_m \sim \pi/4$ in which case $\rho_0
\approx |\sin 2\theta|$.

\item{\bf Near an MSW resonance:} As mentioned above a resonance is only possible for $keV$ sterile
neutrinos for $L\gg T^2/M^2_W$\cite{bell,dibari,kev1,kev2}. For very
small vacuum mixing angle $\sin 2\theta \ll 1$  it proves
illuminating to write the resonance condition $\cos 2\theta =
V_{aa}/\delta M^2$ as  $V_{aa} \sim \delta M^2$ and $\rho_0 \sim
|\sin 2\theta|$, with $V_{aa}$ given by eqn. (\ref{matpotsm}) for
$L\gg T^2/M^2_W$. Therefore $\delta M^2 / k \sim G_F T^3 L $, hence
using eqn. (\ref{smgama}) near the MSW resonance, the ratio \be
\Bigg|\frac{\Gamma_{aa}}{\frac{\delta M^2}{\ook} \rho_0}\Bigg| \sim
\frac{G_F M^2_W}{|\sin 2\theta|} \left( \frac{T^2}{LM^2_W}\right)
\sim \frac{\alpha_w}{|\sin 2\theta|}\left(
\frac{T^2}{LM^2_W}\right)\,. \ee Therefore, the strong damping
condition near the resonance is fulfilled provided  that
$|\sin2\theta|\ll \alpha_w$. With $\alpha_w \sim 10^{-2}$ the region
near an MSW resonance is generally described by the strong damping
regime for $|\sin 2\theta| \lesssim 10^{-3}$, which is likely to be
the case for sterile neutrinos\cite{kev1,kuse2} and is consistent
with constraints from the X-ray background
\cite{hansen,Xray,kou,boyarsky,hansen2}.

 In the resonance region the
sterile production rate is described by the simple rate equation
(see eqn. (\ref{rateNs}) ) \be \dot{N}_s(t) = -\Gamma_2[
N_s(t)-n_{eq}] \label{resrateqn} \ee where the sterile production
rate $\Gamma_2$ is given by eqn. (\ref{Nsrate}) which can be written
as \be \Gamma_2 \sim \sin^2 2\theta \frac{(\delta M^2)^2}{\ook^2
\Gamma_{aa}}\label{resNsrate}\ee and clearly exhibits the
suppression for small vacuum mixing angle and the non-perturbative
nature as a function of $\Gamma_{aa}$.

\end{itemize}

This analysis leads to the conclusion that \emph{away from an MSW
resonance} the weak damping condition holds, sterile neutrino
production \emph{cannot} be described by a simple rate equation but
involves   $\Gamma_{1,2}$ and $\Delta E$. In this regime the quantum
kinetic equations (\ref{dotn11}-\ref{dotn12}) may be
simplified\cite{hoboyste} by neglecting the terms with
$n_{12},n^*_{12}$ in eqns. (\ref{dotn11},\ref{dotn22}) and the terms
with $n_{11}-n(\omega_1);n_{22}-n(\omega_2)$ in eqn. (\ref{dotn12}).
The resulting equations are very simple and their solutions feature
the two  damping rates $\Gamma_1 = \Gamma_{aa}\cos^2\tm;\Gamma_2 =
\Gamma_{aa}\sin^2\tm$. This simplification also holds if the lepton
asymmetry is of the same order of the baryon asymmetry $L \sim
10^{-9}$ in which case $L\ll T^2/M^2$ for $T \gtrsim
3~\textrm{MeV}$\cite{boyhohec,dolgovrev} and no MSW resonance is
available\cite{bell,notzold,dibari}. Near an MSW resonance for
sterile neutrinos with $\sim \textrm{keV}$ mass and $\sin2\theta
\lesssim 10^{-3}$ the strong damping condition holds and $N_s(t)$
obeys a simple rate equation, but the sterile production rate is
\emph{suppressed} by the quantum Zeno effect.

  For $\textrm{keV}$ sterile
neutrinos with small mixing angle $\sin 2\theta \lesssim 10^{-3}$,
the MSW resonance occurs near the scale of the QCD phase transition
$T \sim 180 ~\textrm{MeV}$\cite{kev1,shapo} with the inherent
uncertainties arising from strong interactions and the rapid change
in the effective number of relativistic degrees of freedom in a
regime in which hadronization becomes important. However, as argued
above, near the MSW resonance the strong damping condition is
fulfilled and quantum Zeno suppression \emph{hinders} the production
of sterile neutrinos. As discussed above the sterile distribution
function obeys a simple rate equation with a production rate given
by eqn. (\ref{Nsrate}) or alternatively (\ref{npg2}) which is
strongly suppressed by the factor $1/\wtg^2 \sim  \sin^2 2\theta
/\alpha^2_w \ll 1$. This suppression of the sterile production rate
makes the production mechanism less efficient near the resonance,
thus relieving  the uncertainties associated with the strong
interactions, although these remain in the non-resonant
scenario\cite{dodelson}.

\section{Conclusions}

The production of a sterile species via active-sterile mixing has
been studied in a simple, exactly solvable model that includes all
the relevant ingredients: active-sterile mixing via an off-diagonal
mass matrix and the coupling of the active species to a bath in
thermal equilibrium. The exact solution of the Heisenberg -Langevin
equations allows to obtain the exact time evolution of the
distribution function for the sterile species and the active-sterile
transition probability. Both are determined by the dispersion
relations and damping rates (widths) of the \emph{two quasiparticle
modes in the medium}. These depend on \be \wtg =
\frac{\Gamma_{aa}}{2\Delta E} \ee where $\Gamma_{aa}$ is the
interaction rate of the active species in the absence of mixing and
$\Delta E$ is the oscillation frequency with corrections from
forward scattering (the index of refraction) but no damping. $\wtg
\ll 1;\wtg \gg 1$ correspond to the weak and strong damping regimes
respectively. In the weak damping case the damping rates are
$\Gamma_1=\Gamma_{aa}\cos^2\tm;\Gamma_2=\Gamma_{aa}\sin^2\tm$   the
active-sterile transition probability is given by eqn.
(\ref{Paswdl}), and the time evolution of the sterile distribution
function is given by eqn. (\ref{wdNs}) for vanishing initial sterile
population, both feature these two scales along with the oscillation
time scale. As a result, the time evolution of the sterile
distribution function does not obey a simple rate equation. These
results confirm those of refs.\cite{hobos,hozeno,hoboyste}. The
exact solution allows the systematic exploration of the strong
damping case for which $\wtg \gg 1$ corresponding  to the situation
in which the   interaction rate in the medium is faster than the
oscillation time scale and the quantum Zeno effect is
present\cite{stodolsky}. In this regime we find that the damping
rates of the quasiparticles are $\Gamma_1=\Gamma_{aa};\Gamma_2 =
\Gamma_{aa}\sin^2 2\tm/4\wtg^2$ where $\tm$ is the mixing angle in
the medium. The active-sterile (generalized) transition probability
is

$$\mathcal{P}_{a\rightarrow s} =
\frac{\sin^22\tm}{4\wtg^2}\left[e^{-\Gamma_1t}+
e^{-\Gamma_2t}-2e^{-\frac{1}{2}(\Gamma_1+\Gamma_2)t}\cos[(\Omega_1-\Omega_2)t\right]
$$

In the strong damping regime the oscillation frequency
$\Omega_1-\Omega_2 \propto \cos2\tm$ \emph{vanishes} at an MSW
resonance and the two quasiparticle states \emph{become degenerate}
leading to a breakdown of adiabaticity. The sterile distribution
function obeys a simple rate equation with a sterile production rate
$\Gamma_2$ strongly suppressed for $\wtg^2 \gg 1$. The suppression
of the active-sterile transition probability and the sterile
production rate, and the vanishing of the oscillation frequency in
the strong damping limit are all consequences of quantum Zeno
suppression. The quantum master equation for the reduced density
matrix is derived and shown to be valid in both limits. From it we
obtain the complete set of quantum kinetic equations that yield the
non-equilibrium evolution of the active and sterile distribution
functions. The complete non-equilibrium time evolution of the active
and sterile distribution functions and the coherences are given by
the set of equations (\ref{dotn11}-\ref{dotn12})  along with the
identifications (\ref{Naoftfin},\ref{Nsoftfin}).  The set of kinetic
equations (\ref{dotn11}-\ref{dotn12})  are shown to be equivalent to
the kinetic equations for the ``polarization vector'' often quoted
in the literature. However, unlike these the set
(\ref{dotn11}-\ref{dotn12}) along with
(\ref{Naoftfin},\ref{Nsoftfin}) yield a complete description of the
non-equilibrium dynamics amenable to a straightforward numerical
analysis, the extrapolation to fermionic degrees of freedom is a
straightforward replacement of the equilibrium distribution
functions by the Fermi-Dirac distributions. Furthermore, the
analysis based on the exact solution and the quantum master equation
yield a wealth of information that cannot be easily gleaned from the
set of kinetic equations, for example the active-sterile transition
probability.

For active neutrinos with standard model interactions it is shown
that the weak damping limit describes   the parameter range
\emph{away} from an MSW resonance and that the strong damping limit
\emph{only} emerges near the resonance for very small vacuum mixing
angle, such that $\sin2\theta \lesssim \alpha_w\sim 10^{-2}$. Such
small value is consistent with constraints from the X-ray
background. This result bears important consequences for
cosmological sterile neutrino production. In the resonant production
mechanism of ref.\cite{dodelson} the production rate peaks at the
MSW resonance, however  our analysis, which   includes consistently
the damping corrections, shows that quantum Zeno suppression
\emph{hinders} the sterile production rate near the resonance. For
$keV$ sterile neutrinos the MSW resonance occurs in a temperature
range too close to the QCD phase transition. Hadronization and
strong interactions lead to substantial uncertainties during this
temperature regime which translate into uncertainties in the
production rate. Quantum Zeno suppression of the production rate in
this regime relieves these uncertainties.

\textbf{In summary:} The   set of kinetic equations
(\ref{dotn11}-\ref{dotn12}) (with Fermi-Dirac equilibrium
distributions)  along with the relations
(\ref{Naoftfin},\ref{Nsoftfin}) yield a complete description of the
non-equilibrium dynamics of active and sterile neutrino production
valid in the weak and strong damping limits. Quantum Zeno
suppression is operative near an MSW resonance and suppresses the
sterile production rate, thus relieving potential uncertainties
associated with the QCD phase transition for $keV$ neutrinos.

 \acknowledgements The author thanks C.-M.Ho for fruitful discussions and    acknowledges support
 from the U.S. National Science Foundation through grant award
 PHY-0553418.

\end{document}